\documentclass[10pt]{iopart}

\usepackage{iopams}  
\usepackage{siunitx}
\usepackage{chemformula}  
\usepackage{graphicx}
\graphicspath{{pics/}}
\usepackage{harvard}
\usepackage{hyperref}
\hypersetup{
  colorlinks   = true, 
  urlcolor     = blue, 
  linkcolor    = red, 
  citecolor   = darkG 
}

\begin{document}

\title[Radiative transfer simulation in-situ diagnostics] {Radiative transfer simulations for in-situ particle size diagnostic in reactive, particle growing plasmas\footnote{This is the version of the article before peer review or editing, as submitted by an author to Journal of Physics D: Applied Physics. IOP Publishing Ltd is not responsible for any errors or omissions in this version of the manuscript or any version derived from it. The Version of Record is available online at \url{https://doi.org/10.1088/1361-6463/ac74f6}}}

\author{Julia Kobus$^1$, Andreas Petersen$^2$, Franko Greiner$^{2,3}$ and Sebastian Wolf$^{1,3}$}

\address{$^1$ Institute of Theoretical Physics and Astrophysics, Kiel University, 24118 Kiel, Germany}
\address{$^2$ Institute of Experimental and Applied Physics, Kiel University, 24118 Kiel, Germany}
\address{$^3$ Kiel Nano, Surface and Interface Science KiNSIS, Kiel University, Germany}
\ead{jkobus@astrophysik.uni-kiel.de}
\vspace{10pt}
\begin{indented}
\item[]\today
\end{indented}

\begin{abstract}
When considering particles produced in reactive plasmas, their basic properties, such as refractive index and grain size often need to be known. They can be constrained both ex-situ, e.g., by microscopy, and in-situ by polarimetry, i.e., analyzing the polarization state of scattered light. Polarimetry has the advantage of temporal resolution and real-time measurement, but the analysis is often limited by the assumption of single scattering and thus optically thin dust clouds. This limits the investigation of the growth process typically to grain sizes smaller than about \SI{200}{\nano\meter}. Using 3D polarized radiative transfer simulations, however, it is possible to consider multiple scattering and to analyze the properties of dense particle clouds.  

We study the impact of various properties of dust clouds on the scattering polarization, namely the optical depth of the cloud, the spatial density distribution of the particles, their refractive index as well as the particle size dispersion. We find that ambiguities can occur regarding optical depth and spatial density distribution as well as regarding refractive index and particle size dispersion. Determining the refractive index correctly is especially important as it has a strong impact on the derived particle sizes. With this knowledge, we are able to design an in-situ diagnostics strategy for the investigation of the particle growth process based on radiative transfer simulations which are used to model the polarization over the whole growth process. The application of this strategy allows us for the first time to analyze the polarization measured during a growth experiment in a reactive argon-acetylene plasma for particle radii up to \SI{280}{\nano\meter}.
\end{abstract}


\maketitle
\ioptwocol

\section{Introduction\label{sec:intro}}
In-situ grain size diagnostic using light scattering (ls) techniques are part of dusty plasma physics since its early days in the 1990s.
In 1989 Selywn discovered, that the source of dust particles polluting  microchip production in semiconductor industry is not the wafer itself but the plasma chemistry above the wafer \cite{selwyn_1989_insitu}. Since then the investigation of nanoparticle producing plasmas is part of dusty plasma physics and its technological application \cite{bouchele_book,hollenstein_review2000,watanabe2006,Kortshagen2016,Melino_2021}.

\begin{figure*}[htb]
    \centering
    \includegraphics[width=\linewidth]{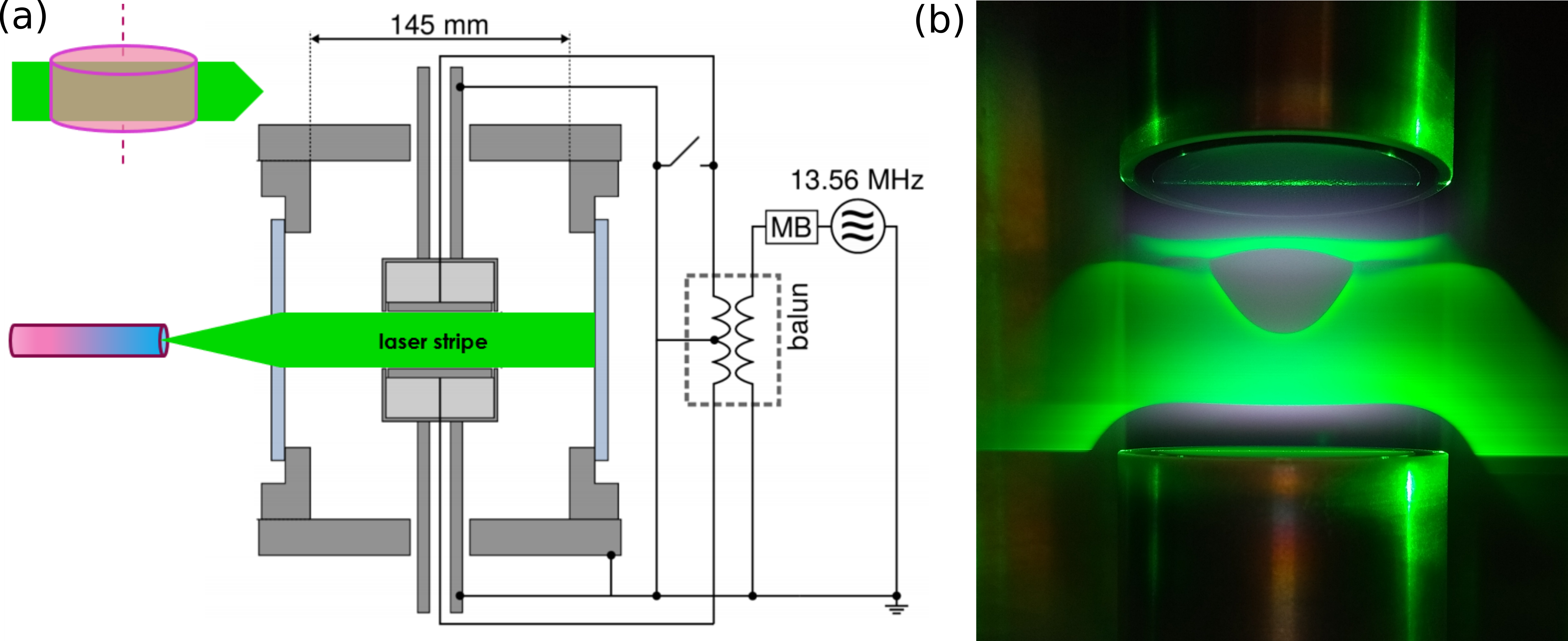}
    \caption{(a) Cross section of a typical cylindrically symmetric plasma chamber used to create plasma-confined nanodust clouds. A laminar flow of argon/acetylene flows through a plasma chamber at a base pressure of around \SI{20}{\pascal}. An rf voltage of $U_{pp}\approx\SI{100}{\volt}$@\SI{13.56}{\mega\hertz} is coupled to shielded planar electrodes via a symmetrizing rf transformer (balun) and drives the plasma. A laser stripe shines through the center of the chamber and the light scattered on the nanoparticles creates a 2D image of the dust cloud. (b) Shows an example cloud, the scattered laser light of a green laser is overlayed with the violetish glow of the argon plasma.
    }
    \label{fig:typchamb}
\end{figure*}

Typical laboratory setups for the investigation of dusty and  reactive plasmas are based on the famous Gaseous Electronics Conference (GEC) cell \cite{Hargis_GEC1994}, which is a low pressure, radio frequency driven, parallel plate reactor.  Figure\,\ref{fig:typchamb}(a) shows a typical setup of a cell, its electrical circuit and a laser stripe, that creates a two-dimensional cross section of the cylindrical dust cloud. Such a cloud may be produced in argon plasma with a acetylene (\ch{C2H2}\;\!) admixture of some percent.
 Figure\,\ref{fig:typchamb}(b) shows the circular electrodes and a nanodust cloud, its shape strongly depends on the electrical confinement of the negatively charged dust and its interplay with the streaming argon ions. In addition to the scattered light of the green laser stripe, the violetish glow of the argon plasma is seen. The creation of a particle cloud starts with a dust free plasma. After the admixture of acetylene the plasma-chemical polymerisation starts, amorphous hydrogenated carbon (\ch{a:C-H}) nanoparticles grow in the plasma (see \citename{greiner_2012_imie} \citeyear*{greiner_2012_imie} and the video in its supplemental material). The time for the growth process to go from molecular precursors to micrometer particles is controlled by the acetylene concentration and takes minutes to hours. It should be emphasized that at ground-based laboratories the particle radius has to be smaller than approximately \SI{400}{\nano\metre} (depending on its mass density and the plasma parameters) to create volume filling clouds, for larger particles gravity pulls the particles down to the lower electrode and only 2D particle layers are created.
 
The characterization of confined dust ensembles needs to answer several questions. First the dust particles need to be constrained in their refractive index and their size.
There are two main approaches: in-situ and ex-situ analysis. Due to its simplicity, microscopy is well-established for ex-situ analysis methods. It was used to give detailed insights into the growth process over time \cite{vandewetering_2016_conclusive}. Although it is possible to extract particles from the dusty plasma without the necessity to stop and restart the growth process and to break the vacuum \cite{dworschak_minimally_2021}, microscopy lacks spatial resolution and cannot be used as a real-time monitor in plasma processing.

For in-situ approaches, however, analysing the light scattered by the growing particles for its polarisation state (polarimetry) has proven effective \cite{hollenstein_1994_diagnostic,hayashi_1994_analysis,groth_2015_kinetic}. Polarimetry offers both, temporal  resolution and real-time capability.
For the standard case of particle growth in reactive  argon-silane (\ch{SiH_4}\;\!) \cite{boufendi_1992} or argon-acetylene (\ch{C_2H_2}\;\!) \cite{denysenko_2006} plasmas, nearly monodisperse and spherical particles are generated having a linear size increase over the growth time. This makes data analysis relatively simple, as the Mie solution of the light scattering process can be used to describe the data and the numerical routines needed to analyze the polarimeter data are provided by standard textbooks \cite{bohren_absorption_1983}. Typically rotating compensator polarimeters (RCP) are used to provide the data \cite{hauge1975}. These systems are commercially available. It is important, however, to measure the full polarisation state of the light, including the degree of polarization. Still the data of a single measurement is insufficient, as both size and refractive index are unknown. But as the particles grow over time, a "kinetic" approach solves this issue \cite{groth_2015_kinetic}. 
With the assumption of a constant refractive index over the whole growth process, homogeneous and spherical particles, monodispersity and single scattering, the CRAS-Mie method presented in \citeasnoun{groth_2015_kinetic} delivers both the (constant) refractive index $N$ and the particle radius over time $a(t)$. 

In this work, we investigate, in which way the limitations and restrictions of the polarimeter size diagnostic can be relaxed and more powerful polarimeter schemes, applicable to a wider size regime, can be developed. For single scattering the consideration of a particle size dispersion is straight forward \cite{Gebauer_2003}. In contrast, the consideration of multi-scattering events requires radiative transfer (RT) simulations. RT simulations solve the radiative transfer equation numerically through, for example, Monte-Carlo simulations in which photon packages are followed as they travel through a dust cloud. In these simulations, decisions about the properties of the photon packages and their interaction with the dust, e.g. scattering events, are made stochastically on the basis of probability distributions given by the properties of the radiation source and the dust particles. \citeasnoun{kirchschlager_-situ_2017} showed that RT simulations can be used to describe the polarimeter data for situations at high dust density. A map of simulated polarimeter curves (providing the relation of the polarimetric angles $\Psi$ and $\Delta$) at different dust densities makes it possible to determine the particle size and the particle density from an experimental $\Delta(\Psi)$-curve. However, the study was restricted to a cylindrical dust cloud at constant density and monodisperse, spherical particles.

In this contribution, we used RT simulations to study the influence of the refractive index, spatial dust density profiles, and a polydisperse size distribution on the polarimeter data.  Knowing the  influence of the these parameters on the shape of the $\Delta(\Psi)$-curve, we developed a data analysis procedure to extract the refractive index and the particle radius over time $a(t)$ from the polarimeter data. The ultimate goal is, to model $\Delta(\Psi)$ over the whole growth process. The application of the new diagnostic to experimental data gives new insight into the physics of the growth process of nanoparticles in a reactive plasma. The analysis of the $\Delta(\Psi)$-curve for particle radii up to \SI{280}{\nano\metre} becomes possible for the first time.

This contribution is structured as follows. First, we describe our approach in Sect.\,\ref{sec:proc}. In the framework of a parameter study presented in Sect.\,\ref{sec:ParameterStudy}, we investigate the impact of the properties of the dust cloud on the polarization of the scattered light. Subsequently, on this basis, we derive implications for the particle growth diagnostics in Sect.\,\ref{sec:implications}. In Sect.\,\ref{sec:diagnostic}, we use the knowledge gained regarding the relevance of the physical properties of the particle cloud to the interpretation of the measured polarization signal to develop an in-situ diagnostic strategy. Finally, in Sect.\,\ref{sec:appl} we apply this strategy to investigate the dust grain growth during an experiment in an rf driven low pressure argon-acetylene plasma.

\section{Procedure}\label{sec:proc}

\subsection{Experimental design} \label{sec:Exp}
The basis for this study is a particle growth experiment in a reactive argon-acetylene plasma. The plasma is created between two parallel electrodes using an rf source of $\approx$\,\SI{8}{\W} at \SI{13.56}{\mega\Hz} for a typical argon pressure of \SI{21}{\pascal}. The two electrodes with a diameter of \SI{6}{\centi\meter} have a distance of \SI{3}{\centi\meter}. By feeding acetylene,   \ch{a:C-H} particles grow in the argon plasma up to a radius of a few hundred nanometers until the particles leave the plasma due to the increase of ion drag force and gravitational force. 

The resulting dust cloud is probed by means of Mie-scattering polarimetry. For this purpose, the polarization state of the scattered light of a red ($\lambda = \SI{662.6}{\nano\meter}$) laser is measured. The laser beam is arranged in such a way that it passes the cloud at its outer region with an offset of \SI{2.8}{\centi\meter} in the direction of the orthogonally positioned polarimeter. This minimizes both the optical depth and thus multiple scattering along the path to the polarimeter as well as the impact of a dust-free void. Such a void is located in the center of the cloud throughout the entire course of the experiment. The voidsize increases with the dust size over time and is well below \SI{1}{\centi\meter} for the duration of the experiment. The light of the laser beam is polarized linearly, oriented \SI{45}{\degree} with respect to the scattering plane, which is determined by the incident laser beam ($x$-direction) and the $y$-axis. The latter corresponds to the line connecting the beam and the polarimeter. The polarimeter is then used to measure the components  of the Stokes vector $S = \left(I,Q,U,V\right)^T$ \cite{collett_polarized_2012} and thus the polarization state of the scattered light. Here, the (total) intensity $I = I_\mathrm{p} + I_\mathrm{u}$ is composed of the polarized fraction $I_\mathrm{p}$ and the unpolarized fraction $I_\mathrm{u}$. Accordingly, we normalize the Stokes vector based on the polarized intensity $I_\mathrm{p} = \sqrt{Q^2 + U^2 + V^2}$:
\begin{equation}
    \left(\begin{array}{c} q \\ u \\ v \end{array}\right) = \frac{1}{\sqrt{Q^2 + U^2 + V^2}} \left(\begin{array}{c} Q \\ U \\ V \end{array}\right), 
\end{equation} 
where the normalized Stokes parameters $q$, $u$, and $v$ are functions of the time-dependent physical properties of the dust cloud, and thus functions of the time-dependent particle size. 
Traditionally, Mie polarimetry (also called Mie ellipsometry) uses $\Delta(\Psi)$-curves for data analysis.
Thus, we calculate the polarimetric angles
\begin{equation}
    \Psi = \frac{1}{2} \arccos\left(-q\right) 
\end{equation}
and
\begin{equation}
    \Delta = \arctan\left(-\frac{v}{u}\right).
\end{equation}

\subsection[The radiative transfer code POLARIS]{The radiative transfer code POLARIS\footnote{While the pioneering study \cite{kirchschlager_-situ_2017}  was performed with \texttt{Mol3D} \cite{ober_tracing_2015}, for this study we use the more versatile radiative transfer software \texttt{POLARIS} because of its better flexibility in terms of already implemented radiation sources and possible model geometries. Comparing the results in \citeasnoun{kirchschlager_-situ_2017} with results of corresponding \texttt{POLARIS} simulations, we find no significant deviations.}} \label{sec:RT}
To model the Mie scattering of laser light in a dust cloud we perform RT simulations with the 3D Monte-Carlo continuum RT code \texttt{POLARIS}\footnote{Available at \protect\url{http://www1.astrophysik.uni-kiel.de/~polaris/}} \cite{reissl_radiative_2016}. The code solves the equation of radiative transfer numerically applying the Monte-Carlo method. \texttt{POLARIS} was initially designed to simulate the observational appearance of astronomical objects containing dust particles of nanometer to millimeter size. The ability to simulate the interaction of radiation with an optically thick cloud of dust grains, however, makes \texttt{POLARIS} as well suited for our study. In the following, those aspects of the scattering simulations which are required for our study are summarized in brief.

First, a given number $N_p$ of photon packages is generated. These packages are characterized by their traveling direction, wavelength $\lambda$ and the Stokes vector $S = \left(I,Q,U,V\right)^T$, where the power $L$ of the radiation source is distributed equally among all photon packages. Then, the path of the photon packages through a three-dimensional grid  is followed. The grid contains the physical properties of the dust clouds, namely its particle number density distribution as well as the extinction cross section, albedo and the scattering matrix (Mueller matrix) of a given dust grain. These parameters characterize the interaction of the laser light with the dust. The optical depth $\tau$, and thus the distance a photon package travels to the next scattering event, is obtained from a random number $z \in [0,1)$ following the expression
\begin{equation}
    \tau = - \ln(1-z).
\end{equation}
When it comes to a scattering event, the photon package receives a new travel direction, which is drawn according to the scattering angle distribution given by the optical properties of the dust, and the Stokes vector is modified according to the Mueller matrix $M$:
\begin{equation}
 S' = M S.
\end{equation}
The random walk of the photon packages is followed until they leave the grid and eventually reach the simulated detector.

For improving the signal-to-noise ratio, two optimization methods are used. The enforced-first-scattering method \cite{cashwell_practical_1959} is applied to improve the photon statistics at low optical depths where scattering is unlikely. For this purpose, every photon package is split in two new packages, whereof one is forced to scatter inside the grid. Here, the energy is distributed among both photon packages in such a way that the low probability of the forced scattering event is compensated. Another issue affecting photon statistics stems from the fact that the detector covers only a small solid angle. Thus, most photon packages do not leave the model space in the direction of the detector. In order to overcome this problem, the peel-off technique is used \cite{yusef-zadeh_bipolar_1984}. Here, at each scattering event a peel-off photon package is generated that is sent directly to the detector. The peel-off photon package obtains a part of the energy $E_\mathrm{ori}$ of the original photon package, which is calculated according to the probability $P(\Omega)$ for the original photon package to be scattered towards the detector and the optical depth $\tau_\mathrm{det}$ between the point of interaction and the detector:
\begin{equation}
    E_\mathrm{peel-off} = E_\mathrm{ori} P(\Omega) \exp(-\tau_\mathrm{det}).
\end{equation}
Here, $\Omega$ is the angle between traveling direction of the original photon package and the direction to the detector.

\subsection{Parameter study: Model description} \label{sec:model}
\begin{figure}
    \centering
    \includegraphics[width=\linewidth]{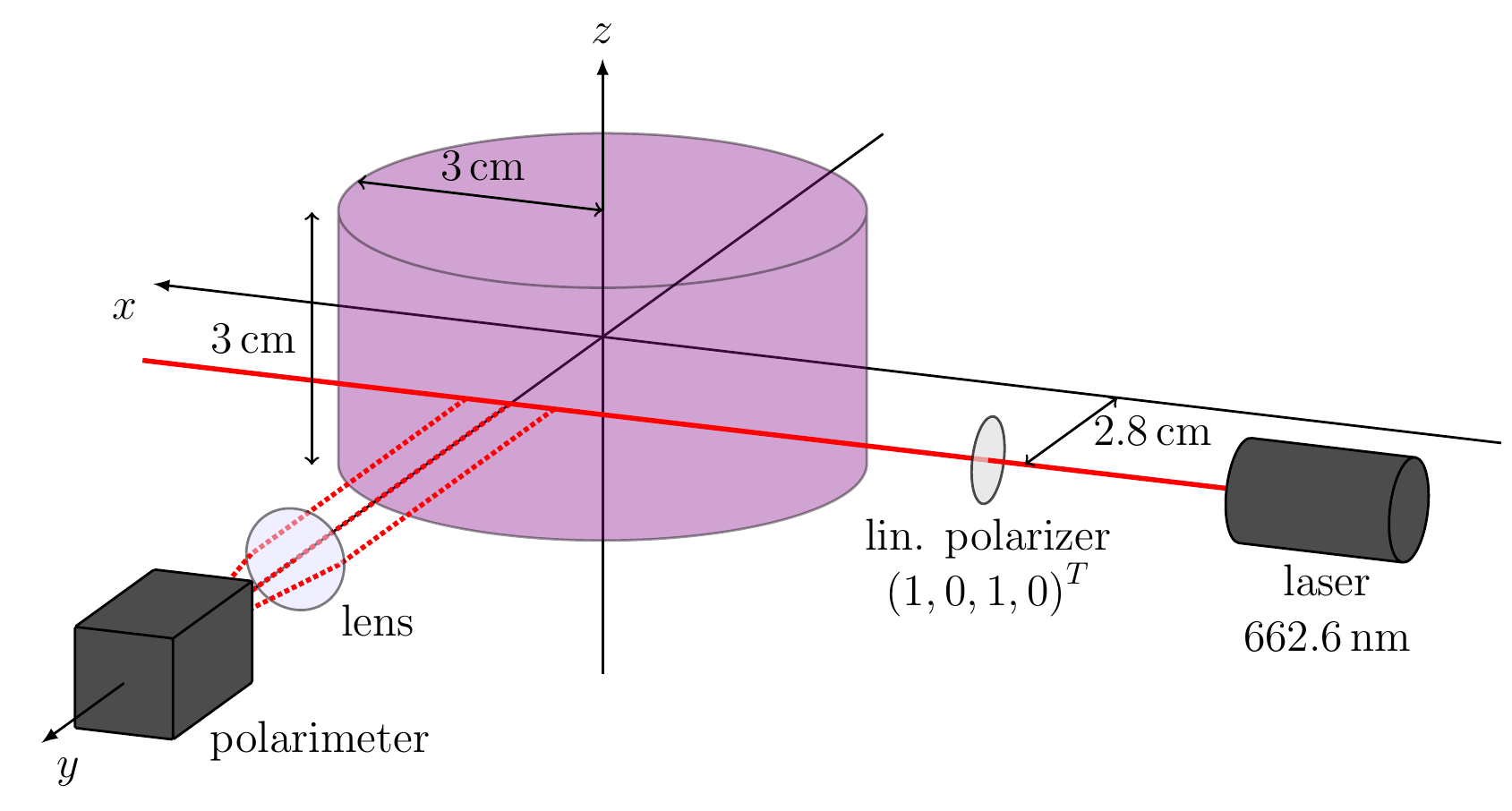}
    \caption{Model setup for the RT simulations (see description in Sect.\,\ref{sec:model}) based on the experimental setup described in Sect.\,\ref{sec:Exp}.}
    \label{fig:experiment}
\end{figure}

The reference model described below and illustrated in figure\,\ref{fig:experiment} serves as the basis for the parameter study to investigate the impact of various quantities on the scattering polarization. As in the experimental setup, the radiation source is a red laser beam ($\lambda = \SI{662.6}{\nano\meter}$), which is linearly polarized \SI{45}{\degree} to the scattering plane, described by its Stokes vector $S = \left(1,0,1,0\right)^T$ and positioned with an offset of $\delta y = \SI{2.8}{\centi\meter}$ to the symmetry axis with the photons propagating in x-direction. 

In accordance to the cylindrical geometry of the discharge, we assume a homogeneous cylinder with a height of \SI{3}{\centi\meter} and a diameter of \SI{6}{\centi\meter} for the dust cloud. Within this cylinder, the number density of the dust is constant with $n = \SI{e13}{\per\cubic\meter}$. We assume monodisperse growth \cite{groth_2015_kinetic}, using a single particle radius in the range of \SIrange{20}{300}{\nano\meter} (size step of \SI{5}{\nano\meter}) in the RT simulation for each time step. The refractive index of the dust is chosen to be $N = 1.8 + 0.05i$ which is within the range of refractive indices found for multiple particle growth experiments in an argon-acetylene discharge \cite{groth_2015_kinetic}. From the refractive index, we calculate the extinction cross sections as well as the Mueller matrix with the code \texttt{miex}\footnote{In contrast to the code \texttt{bhmie} from \citeasnoun{bohren_absorption_1983}, \texttt{miex} is suitable for the simulation of Mie scattering in case of arbitrarily large size parameters} \cite{wolf_mie_2004}, which is embedded in \texttt{POLARIS}.

\begin{figure}
    \centering
    \includegraphics[width=\linewidth]{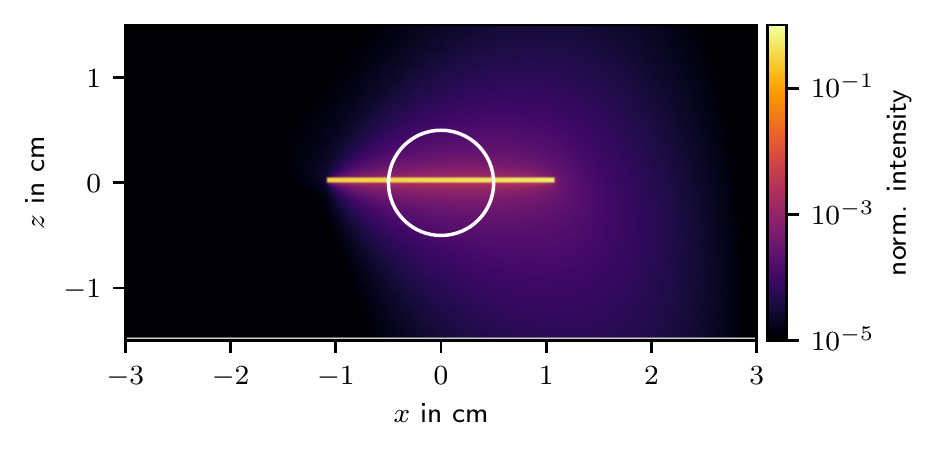}
    \caption{Illustration of the detector image of the normalized total intensity produced by the RT simulations based on the reference model with a particle radius of \SI{200}{\nano\meter}. The white circle indicates the detector area over which the radiation, i.e., the components of the Stokes vector are integrated in accordance with the field of view of the polarimeter.}
    \label{fig:rtdetector}
\end{figure}

The polarimeter is arranged perpendicular to the laser beam in positive y-direction. In the experiment the only photons with parallel trajectories arrive at the polarimeter, as a telecentric lens is put in front of the detector. For the RT simulations, the actual distance can be ignored, as we are solely interested in the normalized Stokes parameters.  
The field of view of the polarimeter covers a radius of \SI{5}{\milli\meter} of the center of the dust cloud. Accordingly, we integrate the radiation over the innermost region of the detector image obtained in the RT simulations before normalizing the Stokes parameters (see figure\,\ref{fig:rtdetector} for illustration).

Based on the described reference model, we vary the following parameters in Sect.\,\ref{sec:ParameterStudy} to investigate the impact of different physical properties of the dust cloud on the polarization state of the scattered light:
\begin{description}
\item[Optical depth] At high optical depth, multiple scattering can no longer be neglected and influences the polarization state of the detected scattered light. To investigate this impact, we vary the optical depth by scaling the constant density \mbox{$n = \alpha \times \SI{e13}{\per\cubic\meter}$} within the cylinder by a factor $\alpha \in [0.1,10]$. A value of $\alpha = 1$ results in a typical dust density from experimental observations \cite{tadsen_2015}.
\item[Dust density distribution] A cylindrical dust cloud with homogeneous density is only a very simple approximation of the dust distribution present in the experiment. We investigate the impact of the dust density distribution to determine whether the true density distribution is needed for the analysis of the particle growth. For this purpose, we use a trapezoidal density distribution as motivated by experimental observations \cite{tadsen_2015}. This density distribution is constant inside the radius $r_\mathrm{const}$ with \mbox{$n(r\leq r_\mathrm{const}) = \alpha \times \SI{e13}{\per\cubic\meter}$} and then decreases linearly to zero up to the outer radius $r_\mathrm{out}$. The plateau radius $r_\mathrm{const}$ is varied between \SIrange{0}{3}{\centi\meter}. The respective outer radius is selected in such a way that the optical depth along the laser beam remains the same as for the reference case of a cylindrical cloud with constant density. The resulting density distributions as well as the particle density along the laser beam are illustrated in figure\,\ref{fig:imp_dist}.
\item[Refractive index] Since the refractive index of dust particles is not necessarily the same as that of a solid of the same material, the refractive index has to be determined as part of the diagnostics based on the measured scattered light. Thus, we investigate its impact by varying both the real ($m_\mathrm{r} \in \left[1.5,2.1\right]$) and imaginary part ($m_\mathrm{i} \in \left[0,0.1\right]$) of the refractive index within the range of values for different runs of similar experiments.
\item[Particle size distribution] Both the estimation of the timescales of the processes involved in particle growth and the investigation of particle growth using the imaging Mie technique indicate almost monodisperse growth \cite{greiner_diagnostics_2018}. However, a narrow distribution of particle sizes around the prevailing grain size at a given time during the growth process cannot be neglected. Therefore, we use RT simulations to investigate how the presence of a narrow particle size distribution affects the relation of the polarimetric angles $\Delta(\Psi)$. For this purpose, we choose a log-normal size distribution according to the size distributions found in \citeasnoun{greiner_diagnostics_2018}, considering two widths given by standard deviations of $\sigma = \SI{5}{\percent}$ and $\sigma =\SI{10}{\percent}$.
\end{description}

\section{Impact of various physical properties of the dust cloud on $\Delta(\Psi)$} \label{sec:ParameterStudy}
In the following we investigate the impact of the properties of the dust cloud on $\Delta(\Psi)$ using RT simulations as described in Sect.\,\ref{sec:model}.

\subsection{Optical depth} \label{sec:param_tau}
For high optical depths ($\tau \gtrsim 0.1$), multiple scattering contributes significantly to the scattered radiation measured by the polarimeter. Consequently, the CRAS-Mie method cannot be used to study the growth of larger particles in this regime. However, the effect of a high optical depth in the present experimental setup can be predicted with the help of RT simulations. For this purpose, we calculate $\Delta(\Psi)$ for dust clouds with different optical depths, which we achieve by scaling the homogeneous density within the cloud with the parameter $\alpha$.

\begin{figure*}
    \centering
    \includegraphics{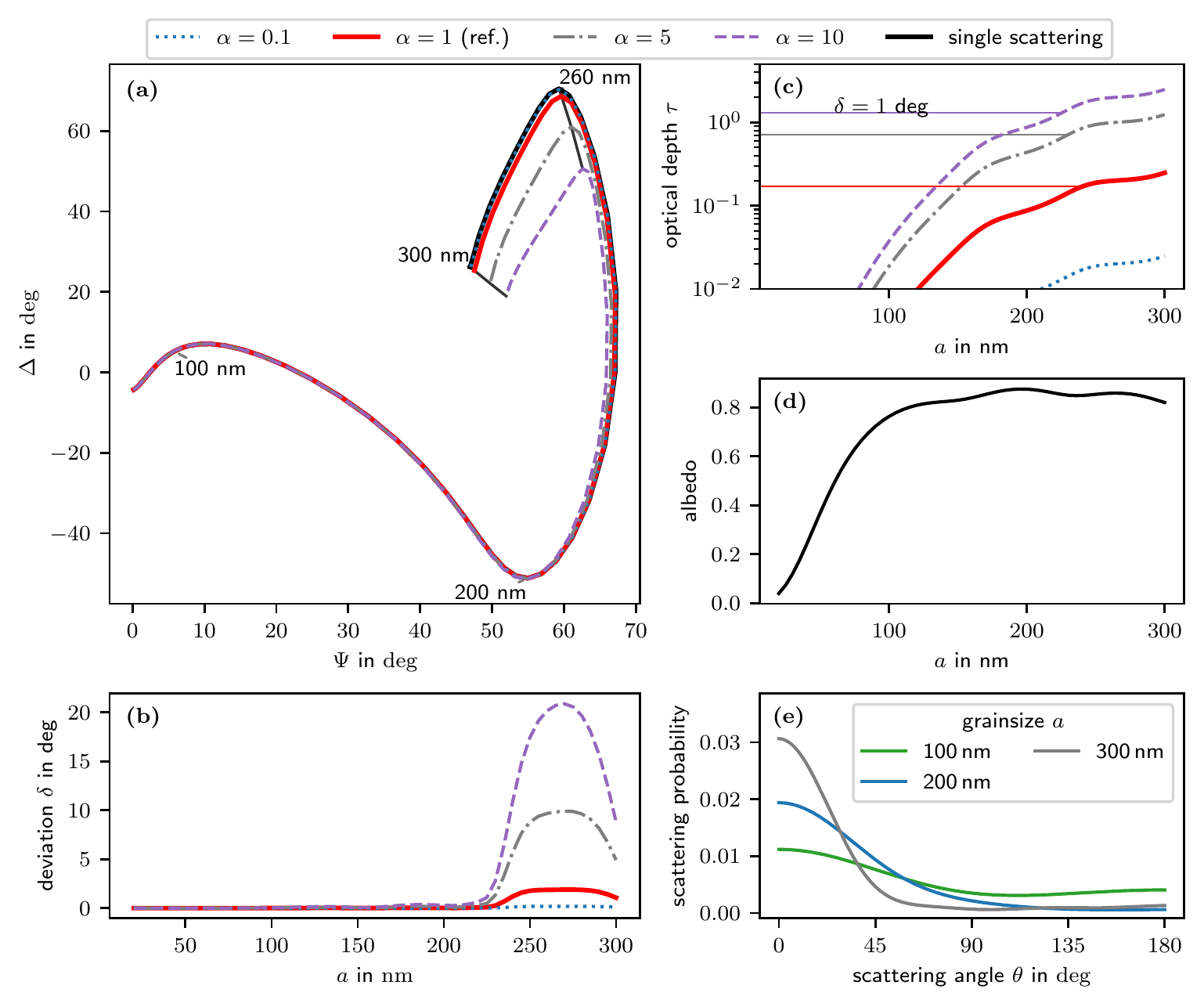}
    \caption{Overview of the impact of the optical depth on the polarization state of the scattered light including the relevant physical quantities for a refractive index $N=1.8+0.05i$. (a) Impact of the dust density scale parameter $\alpha$ on the relation of the polarimetric angles $\Psi$ and $\Delta$. The particle size changes along the $\Delta(\Psi)$-curve, selected particle radii are marked on all curves with black isolines. (b) Deviation of the simulated relation $\Delta(\Psi)$ from $\Delta_\mathrm{s}(\Psi)$ for single scattering with a scattering angle of \SI{90}{\degree} (see equation\,\ref{eq:diff}). (c) Optical depths as a function of particle radius. The horizontal lines mark the optical depths at which $\Delta(\Psi)$ simulated for different density scaling parameters $\alpha$ deviate by more than $\delta>\SI{1}{\degree}$ from the theoretical single scattering result $\Delta_\mathrm{s}(\Psi)$. (d) Albedo as a function of particle radius. (e) Probability distribution of the scattering angles (\SI{0}{\degree}: Forward scattering, \SI{180}{\degree}: Backward scattering) as a function of particle radius.
    }
    \label{fig:imp_alpha}
\end{figure*}

In figure\,\ref{fig:imp_alpha}, we show the resulting relation $\Delta(\Psi)$ together with the deviation from the relation $\Delta_\mathrm{s}(\Psi)$ calculated analytically for single scattering in \SI{90}{\degree}. In addition, we show the physical quantities that in interplay with the optical depth affect the relation $\Delta(\Psi)$.

We find that with increasing values of the density scale parameter $\alpha$ and thus increasing optical depth, the relations $\Delta(\Psi)$ deviate from the single scattering result more and more. In the present configuration, the relations $\Delta(\Psi)$ simulated for higher optical depth and thus increasing impact of multiple scattering all lie within the curve $\Delta_\mathrm{s}(\Psi)$ for single scattering. This characteristic behavior was also found by \citeasnoun{kirchschlager_-situ_2017} for a smaller refractive index of $N = 1.54 + 0.02i$. Above a particle radius of about \SI{200}{\nano\meter} the curves start to deviate visibly from $\Delta_\mathrm{s}(\Psi)$ showing significant deviations above a particle radius of about \SI{225}{\nano\meter}, with maximum deviations close to the maxima.

Noticeably, the deviations from single scattering do not depend on the optical depth alone. To illustrate this, we show the deviation parameter $\delta$ in figure\,\ref{fig:imp_alpha}(b). This quantity is calculated using the Euclidean difference 
\begin{equation} \label{eq:diff}
  \delta = \sqrt{(\Psi_\mathrm{s} - \Psi_\alpha)^2 + (\Delta_\mathrm{s} - \Delta_\alpha)^2}
\end{equation}
between the relation $\Delta(\Psi)$ of models with different values of the density scale parameter $\alpha$ and the relation $\Delta_\mathrm{s}(\Psi)$ for single scattering. In addition, in  figure\,\ref{fig:imp_alpha}(c), the particle radius-dependent optical depth along the path of the incident laser beam
\begin{equation}
    \tau = C_\mathrm{ext} \times n \times 2 \sqrt{\left(\SI{3}{\centi\meter}\right)^2-\left(\SI{2.8}{\centi\meter}\right)}, 
\end{equation}
where $C_\mathrm{ext}$ is the extinction cross section, is shown. Here, the particle radii and corresponding optical depths are marked, where the difference from the single scattering curve becomes significant with $\delta>\SI{1}{\degree}$.

In the same figure it can be seen that for the densest model with $\alpha = 10$ significant deviations from single scattering occur from a grain radius of \SI{225}{\nano\meter} and an optical depth of about \num{1. 3}, whereas for the less dense models significant deviations occur at only slightly larger particle radii but significantly smaller optical depths, namely in the case of $\alpha = 5$ for \SI{230}{\nano\meter} and $\tau \approx 0.7$ and in the case of $\alpha = 1$ for \SI{240}{\nano\meter} and $\tau \approx 0.17$. For the model with $\alpha = 0.1$, however, no significant deviations occur. This is because even at the largest grain radius of \SI{300}{\nano\meter} only an optical depth of about \num{0.02} is reached, i.e.\ multiple scattering remains negligible.

The reason for the different optical depths, above which multiple scattering significantly affects the measured scattered light, lies in the further properties that affect the scattered radiation arriving at the polarimeter. First, the albedo, i.e., the fraction of the incident radiation that is scattered by a dust grain, varies with particle size. While small particles predominantly absorb the laser light, the albedo increases up to a grain radius of \SI{100}{\nano\meter}, resulting in a probability for scattering of about \SI{80}{\percent} at \SI{100}{\nano\meter} to \SI{300}{\nano\meter} (see figure\,\ref{fig:imp_alpha}(d)). Second, the probability distribution of the scattering angle $\theta$ changes with particle size (see figure\,\ref{fig:imp_alpha}(e)). Small particles scatter preferentially in the forward direction. However, a significant fraction of the red laser light is also scattered in all other directions including \SI{90}{degree} directly toward the detector. With increasing particle radius, the probability of scattering by \SI{90}{degree} and thus the impact of single scattering directly towards the polarimeter decreases while the influence of multiple scattering on the measured scattered light increases.

\subsection{Dust density distribution} \label{sec:param_den}
Especially at high optical depths, where multiple scattering occurs more often, the spatial distribution of the particles scattering the laser light might be relevant for the measured scattered light. At the same time, however, the impact of multiple scattering complicates the correct determination of the dust density distribution using extinction measurements. Using RT simulations, it is possible to trace the scattering of light in particle clouds with arbitrary density distributions, and thus to study the impact of the spatial dust distribution on the relation $\Delta(\Psi)$. This way, we can assess the importance of knowing the density distribution for analysing particle growth.

\begin{figure*}
\includegraphics{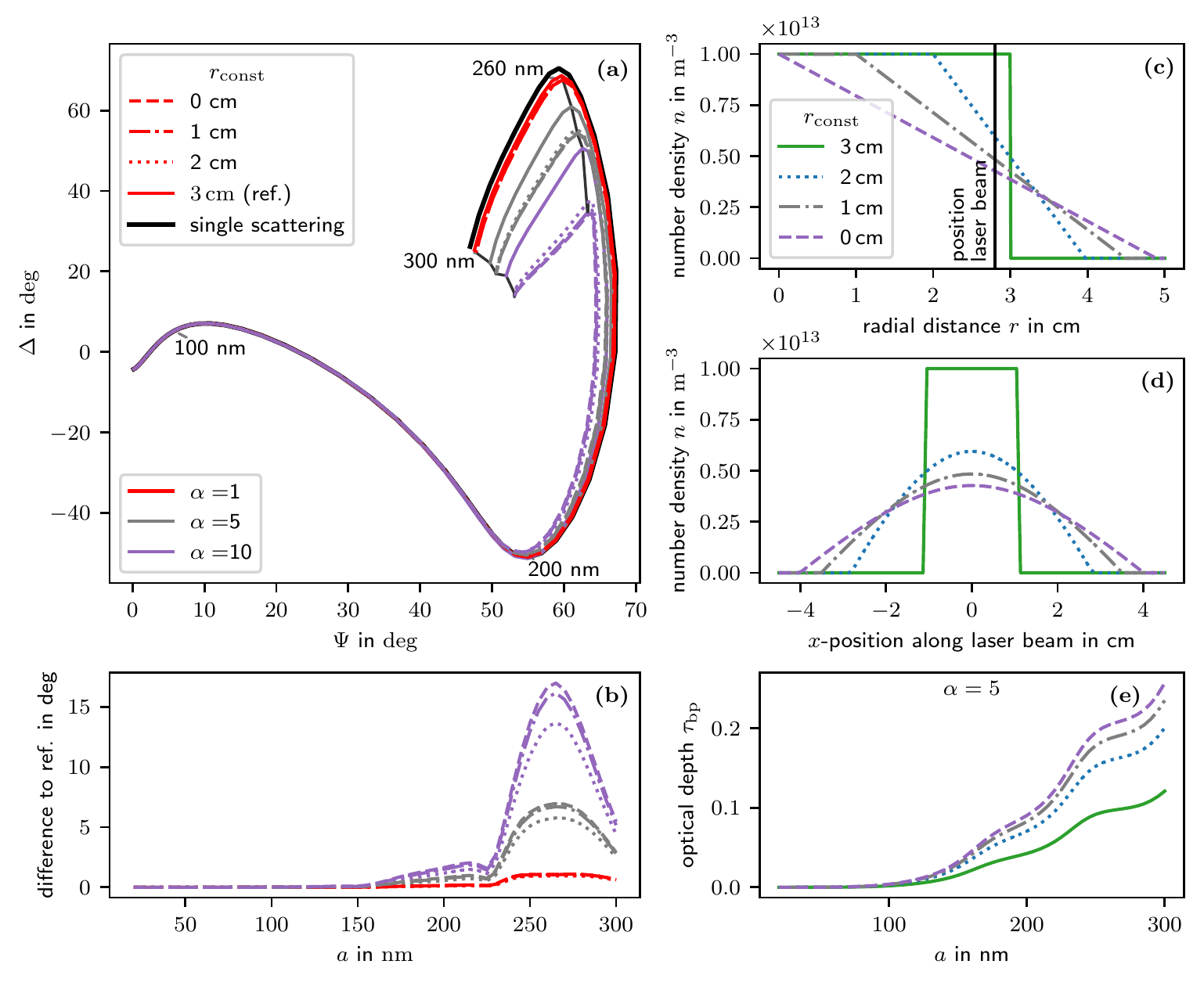}
\caption{Overview of the impact of the spatial particle density distribution on the polarization state of the scattered light including relevant physical quantities. (a) Relation of the polarimetric angles $\Psi$ and $\Delta$ for different density distributions and for assuming single scattering only. (b) Deviation of the $\Delta(\Psi)$-relations of models with different density distributions to the homogeneous reference density distribution with the same optical depth along the laser beam (similar to Eq.\,\ref{eq:diff}). (c) Density distribution of particles for $\alpha = 1$ as a function of radial distance. (d) Particle density for $\alpha = 1$ along the laser beam. As the beam is displaced from the center, the density profile becomes a conical section. (e) Optical depth $\tau_\mathrm{bp}$ between the incident laser beam at (\SI{0}{cm}, \SI{2.8}{cm}, \SI{0}{cm}) and the polarimeter for different density distributions with $\alpha = 5$.
}
\label{fig:imp_dist}
\end{figure*}

Varying the density distribution as described in Sect.\,\ref{sec:model} and illustrated in figure\,\ref{fig:imp_dist}(c,d), while keeping the optical depth constant for each $\alpha$ along the path of the incident laser beam, we obtain the relations $\Delta(\Psi)$ shown in figure\,\ref{fig:imp_dist}(a).

We find discrepancies between the models with different density distributions, which become larger with an increasing value of $\alpha$ and thus increasing optical depth. At $\alpha = 1$, the maximum deviation (in analogy to equation\,\ref{eq:diff}) between the reference model with constant density inside the cylinder ($r_\mathrm{const} = r_\mathrm{out} = \SI{3}{\centi\meter}$) and the model with a linear decrease in density ($r_\mathrm{const} = \SI{0}{\centi\meter}$) is \SI{1.1}{\degree}. Due to the larger value of the density scale parameter of $\alpha = 10$, i.e., at the corresponding higher optical depth, the models differ significantly ($\delta = \SI{17}{\degree}$).

These deviations are due to the fact that while the optical depth along the incident laser beam does not change with the chosen density distributions, the optical depth $\tau_\mathrm{bp}$ between the laser beam and the polarimeter does. Along this path, the optical depth changes the most when going from the constant density model to the model with a trapezoidal density distribution with a constant density up to a distance of \SI{2}{\centi\meter} and a subsequent linear decrease (see bottom right of figure\,\ref{fig:imp_dist}). Accordingly, the largest discrepancies occur between the $\Delta(\Psi)$-relations of the models with constant density and the models with a linear decrease at the outer radii. In addition, at $\alpha = 5$ and $\alpha = 10$ the optical depths between the beam and the polarimeter are just in the range where multiple scattering becomes relevant. In this regard, for the models with smaller $r_\mathrm{const}$, the optical depth is higher and multiple scattering to the polarimeter becomes more important.


\subsection{Refractive index} \label{sec:param_N}
\begin{figure*}
    \centering
    \includegraphics{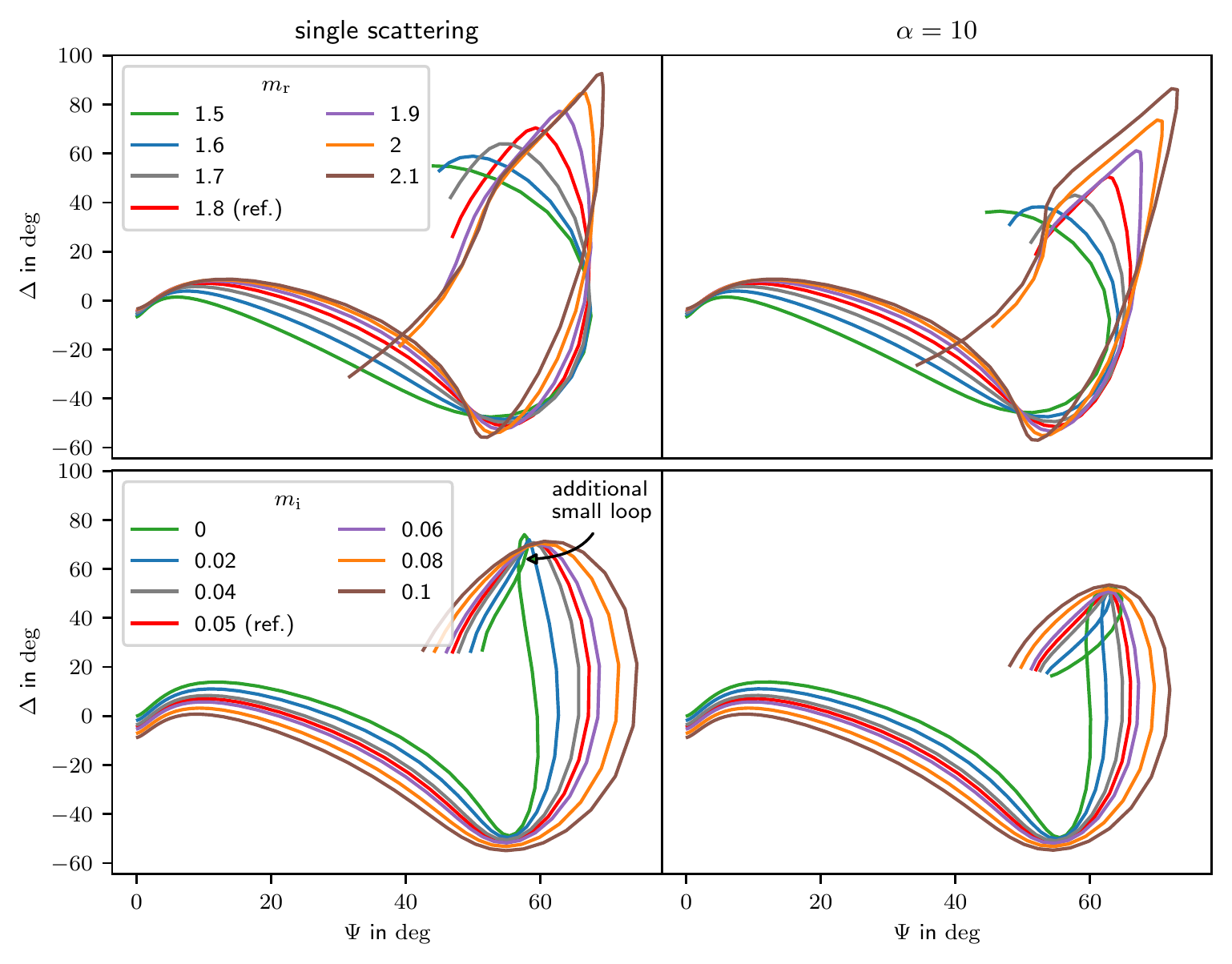}
    \caption{Illustration of the impact of the refractive index on the relation of the polarimetric angles $\Psi$ and $\Delta$. The two rows show the impact of the real (top) and imaginary (bottom) part of the refractive index. The columns indicate the impact of the refractive index for an optically thin model assuming only single scattering (left) and for a model with higher optical depth for a density scale parameter $\alpha = 10$ (right). The red lines in the figures shown in the same column are identical and indicate $\Delta(\Psi)$ for the reference model with $N = 1.8 + 0.05i$.
    }
    \label{fig:imp_refr_idx}
\end{figure*}

The refractive index is one of the free parameters when modeling particle growth using the CRAS-Mie method. Under the assumption of single scattering, the impact of the refractive index can be studied analytically. The investigation of the impact of the refractive index at high optical depths, on the other hand, requires RT simulations.

In figure\,\ref{fig:imp_refr_idx} we show the impact of the real (top) and imaginary part (bottom) of the refractive index on  the relation $\Delta_\mathrm{s}(\Psi)$ of a optically thin model with single scattering only (left) and on $\Delta(\Psi)$ for a model with a density scale parameter $\alpha = 10$ (right). The real and imaginary part of the refractive index have a distinct impact on the shape of the curve. Apart from that, the trends for single scattering and for higher optical depths are rather similar. 

For different real parts of the refractive index, the relations $\Delta(\Psi)$ have a similar starting point at a particle radius of \SI{20}{\nano\meter}. Subsequently, the curves diverge, with higher real part shifting the following local maximum towards larger $\Psi$ and larger $\Delta$. With increasing particle radius, the curves with smaller real part show a rather linear decrease, while the curves with higher real part initially decrease slowly and then more steeply, resulting in a more pronounced minimum with smaller polarimetric angle $\Delta$. The subsequent rounded shape of the curve for small real parts becomes more peaked as the real part increases, with the maximum of $\Delta(\Psi)$ shifting towards larger values of $\Psi$ and $\Delta$ and occurring at smaller particle radii.

Increasing the imaginary part leads to curves with a larger span and a rounder shape. From the smallest particle radius considered (\SI{20}{\nano\meter}) to the minimum, the curves are shifted to smaller values of $\Delta$. Between the minimum of the curve to the largest particle radius (\SI{300}{\nano\meter}), with increasing imaginary part the curve is rounder and broader in $\Psi$-direction. The maximum of the curves, however, is located at about the same position ($\Psi, \Delta$) for the considered range of imaginary parts (\numrange{0}{0.1}). For a given real part of $m_\mathrm{r} = 1.8$, the maximum occurs at a particle radius of \SI{260}{\nano\meter}.
Noticeable is an additional loop, which appears in the case of single scattering for an imaginary part of 0. At higher optical depth with $\alpha = 10$ the loop already appears for an imaginary part of \num{0.02}.

\subsection{Particle size distribution} \label{sec:param_disp}

\begin{figure}
    \centering
    \includegraphics{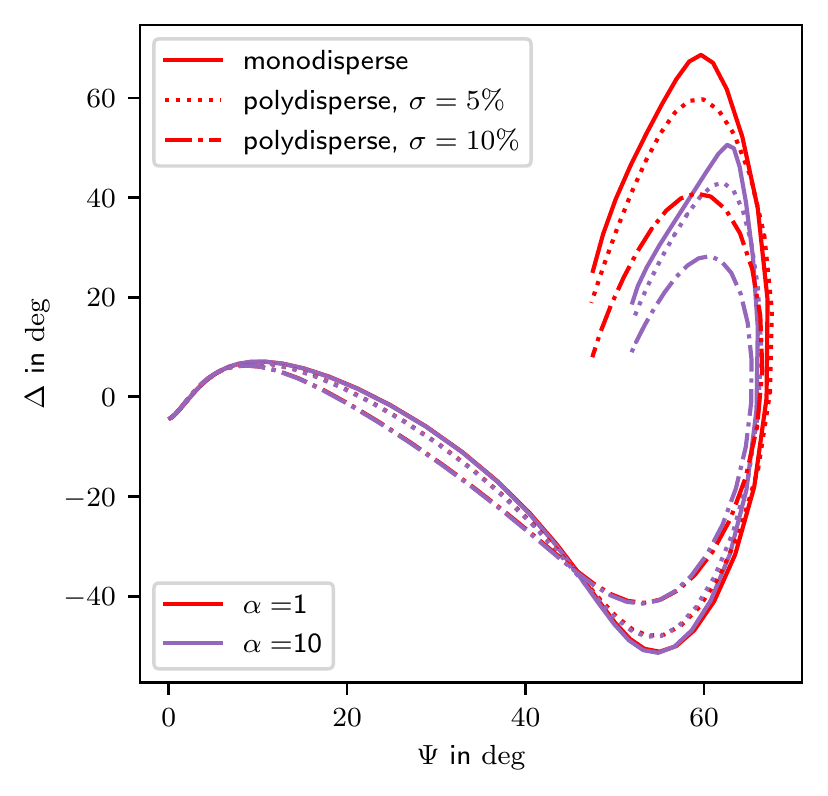}
    \caption{$\Delta(\Psi)$ for models with monodisperse particle growth and models with log-normal particle size distributions for two density scale parameters $\alpha$.}
    \label{fig:imp_sizedist}
\end{figure}

Finally, we investigate the impact of the particle size distribution. For this purpose, $\Delta(\Psi)$-relations for models with monodisperse growth and for polydisperse models with log-normal particle size distributions are shown in figure\,\ref{fig:imp_sizedist}. 

The width of the particle size distribution has a significant impact on $\Delta(\Psi)$. Starting at the first local maximum, the curves for polydisperse growth begin to differ from those for monodisperse growth, with a wider size distribution resulting in a flatter slope. The following minimum of $\Delta(\Psi)$ is less pronounced for a wider size distribution and, subsequently, the curve becomes rounder in the region of large particles. This can be explained by the fact that in the case of the presence of multiple grain sizes, each point $(\Psi, \Delta)$ corresponds to an average of the corresponding grain size range of the $\Delta(\Psi)$-relation of the monodisperse growth. Compared to the monodisperse growth, the $\Delta(\Psi)$-relations therefore appear smeared out, with the effect becoming larger with broader particle size distributions.

Qualitatively, the impact a broader particle size distribution has on the relation $\Delta(\Psi)$ is similar to the impact of a smaller real part of the refractive index. There are clear differences only in the range of large particle sizes at the maximum of the curve. While a smaller real part of the refractive index shifts the maximum to smaller $\Psi$, the change of the $\Psi$-position of the maximum in case of a wide particle size distribution is small. The  particle size distribution primarily varies the $\Delta$-position of the maximum.

\section{Implications for the in-situ diagnostic}
\label{sec:implications}
All investigated physical properties of the dust cloud have a non-negligible impact on the measured polarized scattered radiation. Modeling a measured relation $\Delta(\Psi)$ to investigate the particle growth therefore requires to take all these parameters into account.
In addition, we find the following implications for the in-situ diagnostics of particle growth.

\paragraph{Optical depth:} Our simulations confirm that multiple scattering and thus its impact on the measured polarization becomes relevant above an optical depth of \num{0.1}. However, we also find models in which multiple scattering becomes relevant only at significantly higher optical depths of about 1. Here, the CRAS-Mie method could be used to evaluate $\Delta(\Psi)$ up to significantly larger particle sizes. But the interaction of the various quantities affecting the contribution from multiple scattering is very complex. Thus, it is not straightforward to predict at which optical depth multiple scattering actually contributes significantly. Therefore we advise to apply the CRAS-Mie method to analyze $\Delta(\Psi)$ only up to an optical depth of about 0.1 and to consider RT simulations for higher optical depths. 

\paragraph{Particle distribution:} \label{sec:impl_dist} Above an optical depth of \num{0.1} not only the number of dust grains but also their spatial distribution affects the relation $\Delta(\Psi)$. However, increasing the optical depth along the laser beam and flattening the density distribution have a somewhat similar impact on $\Delta(\Psi)$. Together with the fact that both parameters change in an unknown way during the course of the experiment, this will lead to ambiguities in the subsequent data analysis. As long as we do not aim to determine the particle density distribution in the cloud\footnote{For this purpose, spatially resolved extinction images would be better suited as compared to unresolved measurements of the polarization state of the scattered light.} but intend to investigate the particle growth, this does not pose a problem. In the case that two different models with different combinations of optical depth and density distribution yield similar curves, a given point $(\Psi, \Delta)$ corresponds to the same particle radius. Thus, it is sufficient to find a combination of optical depth and density distribution that can reproduce the measured $\Delta(\Psi)$-relation to correctly determine particle growth.

\paragraph{Refractive index:} Since both real and imaginary parts of the refractive index affect $\Delta(\Psi)$ already at low optical depths from the smallest particle radii, both can be determined without RT modelling using the CRAS-Mie method. Assuming that the refractive index does not change with increasing particle radius, $\Delta(\Psi)$ can also be analyzed at higher optical depths of more than \num{0.1} using RT simulations without having to model the refractive index. 

\begin{figure}
    \centering
    \includegraphics{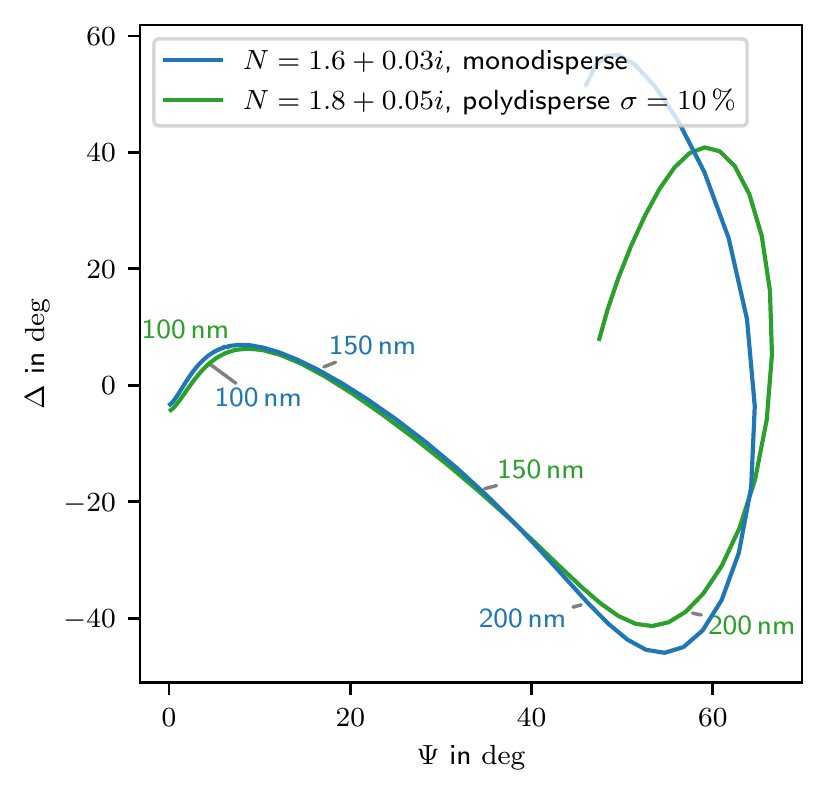}
    \caption{Illustration of the ambiguity in fitting the refractive index and the particle size dispersion. For both relations $\Delta(\Psi)$ corresponding to different refractive indices and different particle size dispersion, the grain radii are marked in the respective color.}
    \label{fig:amb_N_logn}
\end{figure}

\begin{figure}
    \centering
    \includegraphics{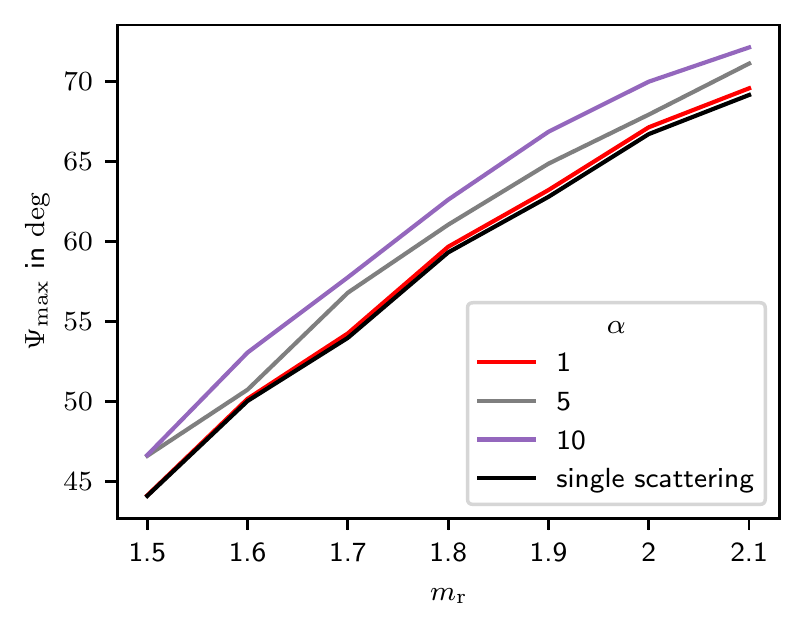}
    \caption{Relation between the $\Psi_\mathrm{max}$ position of the maximum of $\Delta(\Psi)$ and the real part of the refractive index for models with different values of the density scale parameter $\alpha$.}
    \label{fig:psimax}
\end{figure}

However, at small particle sizes a broader particle size distribution has a similar impact on the shape of the $\Delta(\Psi)$-curve as a lower refractive index. In figure\,\ref{fig:amb_N_logn}, the relations $\Delta(\Psi)$ are shown for two exemplary models, one with monodisperse growth and a refractive index of $N = 1.6 + 0.03i$ and the other for polydisperse growth with a log-normal grain size distribution with $\sigma = \SI{10}{\percent}$ and $N = 1.8 + 0.05i$. Up to a particle radius of about \SI{200}{\nano\meter} (referring to $\Delta(\Psi)$ for monodisperse growth with $N = 1.8 + 0.05i$, blue line), both relations $\Delta(\Psi)$ are very similar. Therefore, if only the range of $\Delta(\Psi)$ for small particle sizes is considered for the fitting of the refractive index, e.g. using the CRAS-Mie method, in order to ensure low optical depths, ambiguities may occur with respect to the refractive index and the particle size dispersion. Yet, this refers only to the shape of the $\Delta(\Psi)$-curve. The dependence on the particle radius, is significantly different. That means, for the determination of the particle size, it is important to resolve the ambiguity in the fitting of a measured relation $\Delta(\Psi)$ in the range of low optical depths.

For this purpose, it can be useful to determine the refractive index based on the part of $\Delta(\Psi)$ near the maximum where multiple scattering contributes significantly. Here, the position of the maximum of $\Delta(\Psi)$ provides a good indication of the real part of the refractive index. This is because the real part of the refractive index has by far the largest impact among the studied quantities on the $\Psi_\mathrm{max}$ position of the maximum of $\Delta(\Psi)$. In figure\,\ref{fig:psimax}, the $\Psi_\mathrm{max}$ position of the maximum of $\Delta(\Psi)$ is shown as a function of the real part of the refractive index providing a simple way to estimate the real part for a measured $\Delta(\Psi)$-relation.

\paragraph{Particle size distribution:}
Even narrow distributions of particle size present in the cloud at given times influence $\Delta(\Psi)$ already at low optical depths. Therefore, the size distribution should be taken into account regardless of whether RT simulations or a single scattering approach are used for the modeling. For small imaginary parts of the refractive index, the shape of the maximum of $\Delta(\Psi)$ allows one to easily identify whether mono- or polydisperse growth is present. In particular, for high real parts and small imaginary parts of the refractive index, $\Delta(\Psi)$ shows a tapered maximum in the monodisperse case, which is smeared out in the case of the simultaneous presence of different particle sizes.

\section{Strategy for in-situ particle growth diagnostic} \label{sec:diagnostic}
Based on the known impact of the physical properties of the dust cloud on the relation $\Delta(\Psi)$ and the implications drawn from the parameter study, we propose the following strategy for in-situ diagnostics of particle growth.
\begin{description}
\item[Step 1] Fitting the relation $\Delta(\Psi)$ in one step requires to study a huge parameter space, as the physical properties of the dust cloud, especially the density distribution and the corresponding optical depth may change in unknown fashion during the course of the experiment. However, to limit the parameter space, we can take advantage of the fact that at low optical depths, only the refractive index and the particle size distribution affect $\Delta(\Psi)$. Therefore, we restrict the first modeling step to the estimation of these parameters based on the optically thin part of $\Delta(\Psi)$. This has the additional advantage that no RT simulations are needed for this step, since for single scattering the corresponding relation $\Delta_\mathrm{s}(\Psi)$ can be calculated semi-analytically.
\item[Step 2] For the second step, it is useful to assume that the refractive index and particle size distribution determined in the first step hold for the entire curve or keep changing according to the trends found in the first step. In addition, one can make use of the implication regarding the combination of optical depth along the laser beam and density distribution (see Sect.\,\ref{sec:implications}) and assume a particle distribution leaving a density scale parameter as the only free parameter. But it is not necessary to model the time-dependent density scale parameter. It is sufficient to create simulations for a sufficiently large range of density scale parameters, each of which is constant over the entire curve. The measured $\Delta(\Psi)$-relation with a potentially variable density distribution and optical depth should then lie within the map spanned by these simulated curves.
\item[Step 3] It must be verified that the refractive index and particle size distribution assumed in the second step based on the findings of the first step reproduce the measured $\Delta(\Psi)$ also in the range of high optical depths. If this is not the case, both quantities have to be re-modeled for this range based on RT simulations. 

For this purpose, we suggest the following procedure. We calculate two $\Delta(\Psi)$-relations, one for single scattering and one for a high density scale parameter $\alpha$ resulting in a optical depth exceeding those of the experiment, for each model in the parameter space defined by the refractive index and size distribution. Subsequently, we determine the model that best reproduces the measured $\Delta(\Psi)$-relation by means of least squares. For this purpose, the difference between the model and the measured $\Delta(\Psi)$ is determined as follows: If the measured point is between both simulated curves the difference is zero, otherwise we calculate the shortest distance between the measured point and the nearest of both curves. We then repeat the simulations with different density scale parameters resulting in different optical depths from step 2 with the newly modeled refractive index and particle size distribution.
\item[Step 4] The grain radius corresponding to each measured point $(\Psi, \Delta)$ can be determined as follows. First, the simulated curve nearest to the measured point is identified. Then, the point on the simulated curve closest to the measured point is determined. For this purpose we use the Python package Shapely\footnote{\url{https://pypi.org/project/Shapely/}}. Finally, we determine the grain radius by linear interpolation between the neighboring simulated points $(\Psi, \Delta)$.
\end{description}

\section{Particle growth diagnostic} \label{sec:appl}
\begin{figure}
    \centering
    \includegraphics{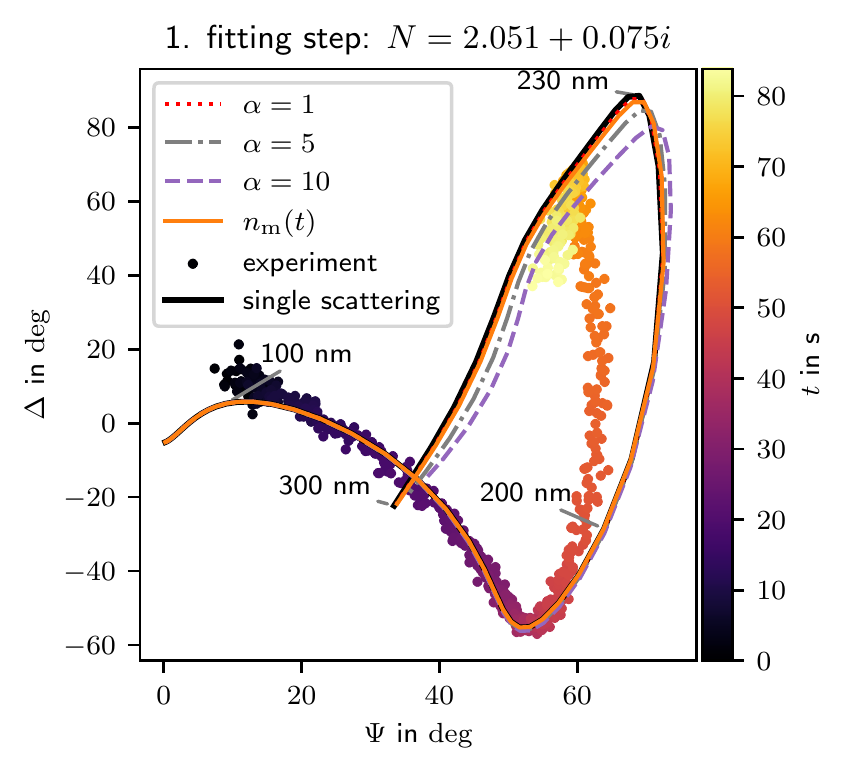}
    \caption{$\Delta(\Psi)$ measured during the experiment described in Sect.\,\ref{sec:Exp} (dots) together with simulated relations $\Delta(\Psi)$ with a refractive index of $N = 2.051 + 0.075i$ and different optical depth.}
    \label{fig:experimental_data}
\end{figure}

Finally, we apply the outlined strategy to study the particle growth during the experiment described in Sect.\,\ref{sec:Exp}.  The $\Delta(\Psi)$-relation obtained during the experiment is shown in figure\,\ref{fig:experimental_data}. 

Following step 1, we first determine the refractive index and the particle size distribution. A prominent feature of the measured curve is the sharply peaked maximum. According to Sect.\,\ref{sec:implications}, we take this as an indication for monodisperse growth. Since the particle size dispersion is not expected to become narrower during the course of the experiment, we assume monodisperse growth for $\Delta(\Psi)$ at every time step. Accordingly, the refractive index remains the only free parameter in this first modelling step. We determine it using the CRAS-Mie method for the part of the relation $\Delta(\Psi)$ measured between \SIrange{10}{35}{\second} (see colorbar in figure\,\ref{fig:experimental_data}) assuming that the refractive index is constant in this range. We obtain a refractive index of $N = 2.051 + 0.075i$.

Following step 2, based on this refractive index and assuming monodisperse growth for the entire curve, we perform RT simulations of models with a constant density distribution $n = \alpha \times \SI{e13}{\per\cubic\meter}$ and different optical depths corresponding to density scale factors $\alpha \in \{1, 5, 10\}$. The resulting $\Delta(\Psi)$-relations are plotted in addition to the measured curve in figure\,\ref{fig:experimental_data}. 

As it can be seen in figure\,\ref{fig:experimental_data}, the simulated curves do not reproduce the measured data for grain radii above about \SI{200}{\nano\meter}. 
This discrepancy is comparable to that found by \citeasnoun{kirchschlager_-situ_2017}, who also discovered that above a grain size of \SI{200}{\nano\meter}, the measured relation $\Delta(\Psi)$ deviates significantly from that for single scattering, whereby the measured curve lies within the curve for single scattering, like in our case. In the case of the significantly lower refractive index ($N = 1.54 + 0.02i$) in \citeasnoun{kirchschlager_-situ_2017}, the measured curve can be reproduced even for larger grain sizes based on RT simulations with a higher optical depth. In our case, however, the increase of the optical depth leads only to small changes of the relation $\Delta(\Psi)$, which do not lead to a better reproduction of the measured curve.

Before re-modeling the refractive index and the grain size distribution as described in step 3, we aim to rule out the possibility that the discrepancies  are caused by an overly simplistic choice of density distribution that cannot be compensated by a variable optical depth (see Sect.\,\ref{sec:implications}). This can be done with the help of extinction images taken during the experiment in addition to the measured polarization state of the scattered light. Assuming that the density distribution in the cloud is cylindrically symmetric, we can estimate it using the Abel inversion procedure. In this way we get the time-dependent density distribution and consequently also the corresponding optical depth.  The relation $\Delta(\Psi)$ provided by the RT simulations for this time-dependent density distribution $n_\mathrm{m}(t)$ is also shown in figure\,\ref{fig:experimental_data}. It turns out that this curve is in the range of those simulated for constant density distributions similar to the curve with $\alpha = 1$. Thus, the discrepancy between the measured and the simulated curves does not result from the choice of the dust density distribution. Instead it is caused by at least one of the values found for the refractive index and the particle size distribution using the CRAS-Mie method not reproducing the properties of the dust cloud at least part of the course of the experiment.

We therefore fit the refractive index and particle size distribution based on RT simulations again, where we consider the entire measured relation $\Delta(\Psi)$ and do not exclude polydisperse growth, at the same time assuming that both quantities are constant over time. For this purpose, we perform RT simulations for the following parameter space:
\begin{itemize}
    \item $m_\mathrm{r} \in \{1.8, 1.9, 2.0, 2.1\}$
    \item $m_\mathrm{i} \in \{0, 0.02, 0.04, 0.06, 0.08, 0.1\}$
    \item monodisperse, polydisperse with $\sigma \in \{0.05, 0.1\}$
\end{itemize}
The bounds of the parameter space regarding the real part of the refractive index are based on the estimates using the CRAS-Mie method ($m_\mathrm{r} = 2.051$) and considering the maximum of $\Delta(\Psi)$ ($m_\mathrm{r} \approx 1.8$, compare figure\,\ref{fig:psimax}). Following step 3, we do not model the density distribution and optical depth but calculate for each model the relation $\Delta_\mathrm{s}(\Psi)$ for single scattering and $\Delta(\Psi)$ for a high optical depth using a density scale factor of $\alpha = 5$. 

\begin{figure}
    \centering
    \includegraphics{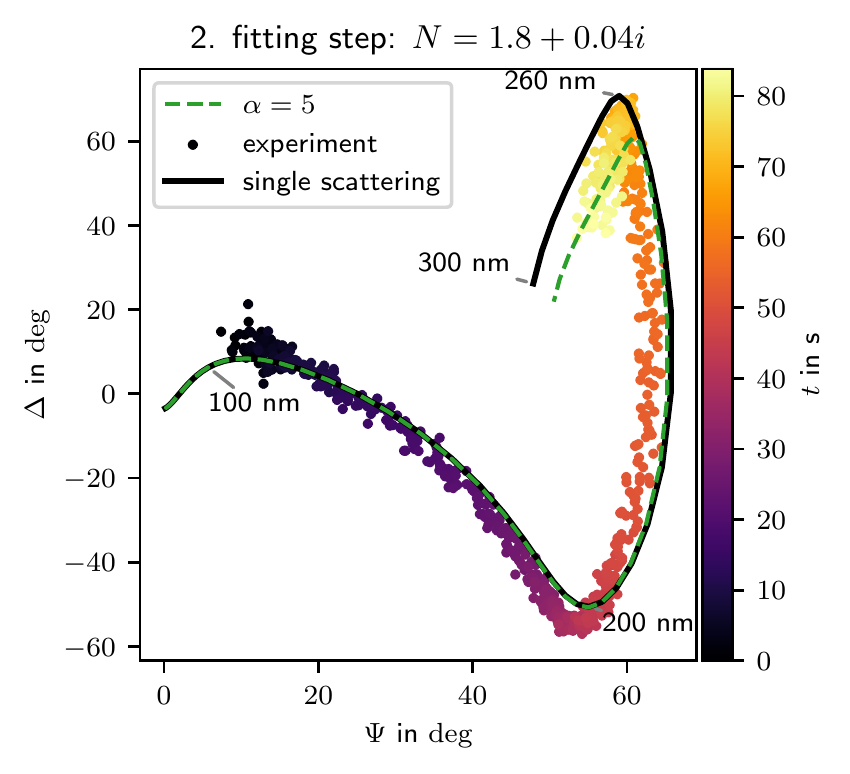}
    \caption{$\Delta(\Psi)$ of the best-fit model considering the full measured relation $\Delta(\Psi)$ ($N = 1.8 + 0.04i$, monodisperse particle growth) and the best-fit model for considering only $(\Psi, \Delta)$-points measured after \SI{45}{\second} ($N = 1.8 + 0.02i$, monodisperse particle growth) simulated for single scattering and a high optical depth corresponding to a density scale factor of $\alpha = 5$.}
    \label{fig:diag2}
\end{figure}

Overall, the measured $\Delta(\Psi)$ is best reproduced by the model with a refractive index of $N = 1.8 + 0.04 i$ and monodisperse growth. The corresponding curves for single scattering and $\alpha = 5$ are shown together with the measured relation $\Delta(\Psi)$ in figure\,\ref{fig:diag2}. It turns out that the model reproduces the measured curve only mediocrely. For small grain radii, where only single scattering occurs, the model overestimates the measured curve in $\Delta$ direction. Especially the pronounced minimum of the measured curve is not reproduced by the model, where the minimum of the model curve is less pronounced and lies at larger $\Psi$ and $\Delta$. For particle radii between \SIrange{200}{260}{\nano\meter} $\Delta(\Psi)$ of the model is too roundish and overestimates the measured curve in $\Psi$ direction. After the maximum, $\Delta(\Psi)$ of the model does not decrease steeply enough. Both could be improved with a smaller imaginary part of the refractive index, but this leads to an even stronger overestimation of the measured $\Delta(\Psi)$ at small particle radii.

\begin{figure}
    \centering
    \includegraphics{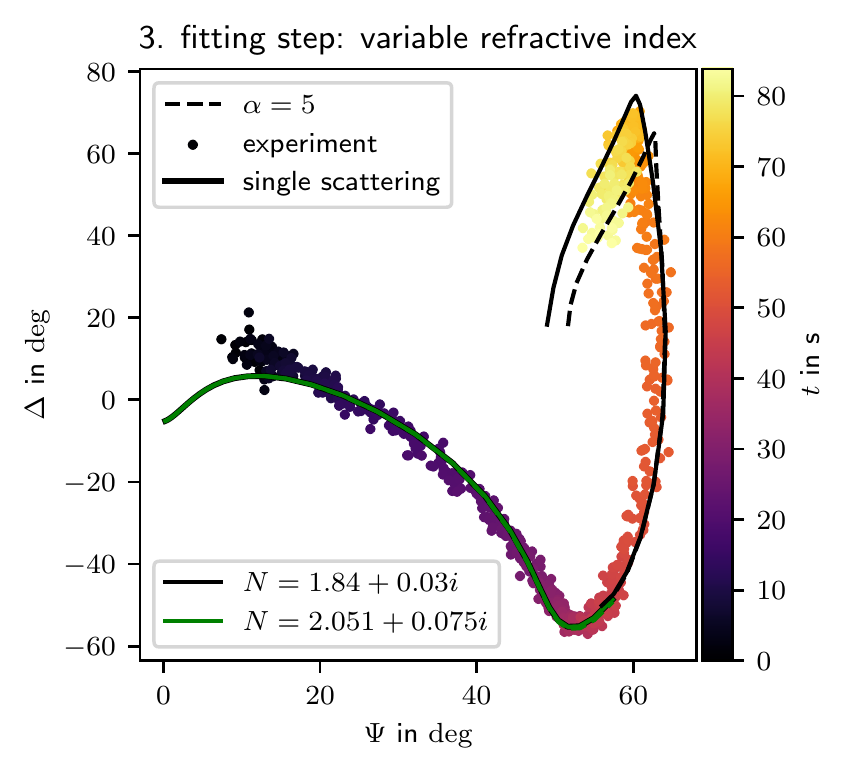}
    \caption{$\Delta(\Psi)$ simulated for a variable refractive index (\SIrange{20}{190}{\nano\meter}: $N = 2.051 + 0.075i$, \SIrange{200}{300}{\nano\meter}: $N = 1.84 + 0.03i$).}
    \label{fig:diag3}
\end{figure}

In order to better reproduce the measured relation $\Delta(\Psi)$ as a basis for the determination of the particle growth, we allow the refractive index of the model to change. Up to $t=\SI{45}{\second}$, the measured curve is reproduced by the refractive index $N = 2.051 + 0.075i$ which we determined using the CRAS-Mie method. The $(\Psi, \Delta)$-points obtained after \SI{45}{\second} are then modeled using RT simulations with higher accuracy in the range of the refractive index estimated for the position of the maximum of $\Delta(\Psi)$s:
\begin{itemize}
    \item $m_\mathrm{r} \in \{1.78, 1.8, 1.82, 1.84, 1.86, 1.88, 1.9\}$,
    \item $m_\mathrm{i} \in \{0, 0.01, 0.02, 0.03\}$.
\end{itemize}
The model with a refractive index of $N = 1.84 + 0.03i$ reproduces the data measured after \SI{45}{\second} best. In figure\,\ref{fig:diag3}, simulated $\Delta(\Psi)$ for $N = 2.051 + 0.075i$ in the particle radius range \SIrange{20}{190}{\nano\meter} and for $N = 1.84 + 0.03i$ in the range \SIrange{200}{300}{\nano\meter} are plotted alongside the measured $\Delta(\Psi)$-relation. With this changing refractive index, the entire measured relation $\Delta(\Psi)$ is well reproduced. 

\begin{figure}
    \centering
    \includegraphics{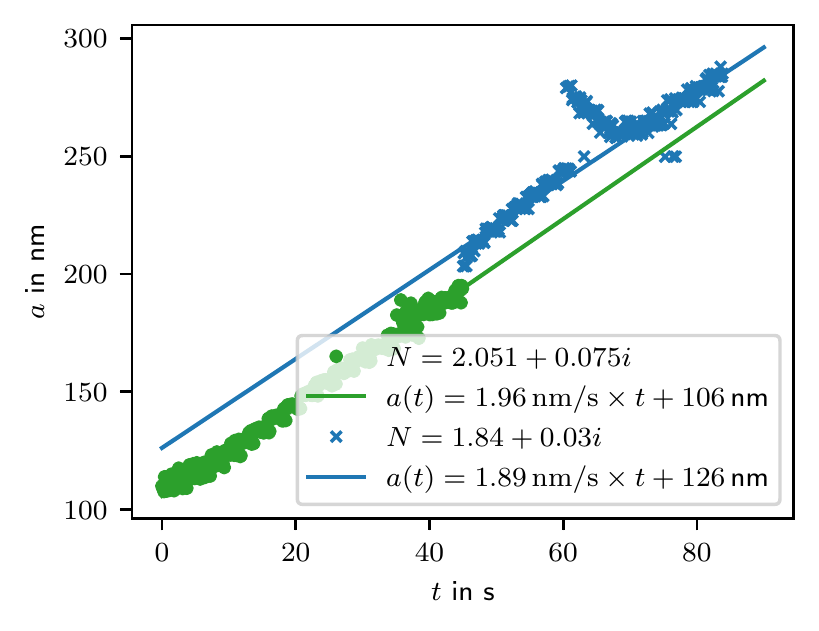}
    \caption{Particle radius as a function of time. The particle radii determined according to step 4 for each measured $(\Psi, \Delta)$-point are plotted as circles (\SIrange{20}{190}{\nano\meter}: $N = 2.051 + 0.075i$) and as crosses (\SIrange{200}{300}{\nano\meter}: $N = 1.84 + 0.03i$). In addition, linear fits (solid lines) for the derived particle growth are shown for both parts of the relation $\Delta(\Psi)$ modeled with different refractive indices.
    }
    \label{fig:at}
\end{figure}
Finally, we can now determine the particle growth following step 4. The resulting particle radii are plotted against time in figure\,\ref{fig:at}. For the period up to \SI{45}{\second} \mbox{($N = 2.051 + 0.075i$)}, the dust grains grow linearly from \SIrange{105}{195}{\nano\meter}. A linear fit results in a growth rate of \SI{1.96}{\nano\meter\per\second}. After \SI{45}{\second} ($N = 1.84 + 0.03i$) the particle radius increases rapidly by about \SI{17}{\nano\meter}. Afterwards, the particle radii increase linearly again. Between \SIrange{60}{70}{\second}, there is an additional sudden increase in particle size, which then drops again. However, this is not physical but due to the fact that the determination of the particle radius in the region of the sharply peaked maximum is inaccurate. After \SI{70}{\second}, the particles grow linearly again up to a radius of about \SI{285}{\nano\meter}. A linear fit, where we do not consider data points measured between \SIrange{60}{70}{\second}, yields a growth rate of \SI{1.89}{\nano\meter\per\second}. Thus, after the sudden increase in particle size at the change of refractive index, the particles continue to grow at about the same growth rate as before.

The linear growth with a growth rate of \SI{1.96}{\nano\meter\per\second} up to a particle radius of \SI{195}{\nano\meter} is consistent with the growth rates of similar growth experiments in argon-acetylene discharges found for particle sizes $\lesssim$\SI{200}{\nano\meter} \cite{boufendi_1992,denysenko_2006,groth_2015_kinetic}. Interestingly, there is also some indication that the sudden increase in grain size at about \SI{200}{\nano\meter} is not an exceptional phenomenon. \citeasnoun{dworschak_minimally_2021} used a minimally invasive method to extract dust grains from an argon-acetylene discharge during the course of a growth experiment and were thus able to analyze the particle size ex-situ. The particle sizes measured in this way indicate a sudden increase of the particle size between about \SIrange{200}{250}{\nano\meter}  \citeaffixed{dworschak_minimally_2021}{see figure 10 in}. 

\section{Conclusion} \label{sec:conclusion}
3D radiative transfer simulations enable the in-situ analysis of optically thick dust clouds based on measurements of the polarization state of scattered light. We aimed at developing a diagnostic strategy for the investigation of the particle growth process in a nanoparticle producing plasma that utilizes radiative transfer simulations to model the relation $\Delta(\Psi)$ measurable during the growth process.

For this purpose, in Sect.\,\ref{sec:ParameterStudy} we first investigated the impact of the properties of the particle cloud (optical depth: Sect.\,\ref{sec:param_tau}, spatial distribution of the dust particles: Sect.\,\ref{sec:param_den}, refractive index: Sect.\,\ref{sec:param_N}, and particle size dispersion: Sect.\,\ref{sec:param_disp}), in order to evaluate on which variables the planned in-situ diagnostic is sensitive to and where ambiguities may occur.

Based on this parameter study, we can conclude the following implications for the in-situ diagnostics (details in Sect.\,\ref{sec:implications}). The results of our radiative transfer simulations support the rule of thumb that above an optical depth of \num{0.1} multiple scattering can have a significant impact on $\Delta(\Psi)$ \cite{Hulst57}. Taking multiple scattering into account is possible with RT simulations, whereas analysis strategies based on single scattering assumptions, such as the CRAS-Mie method, should not be used above an optical depth of \num{0.1}. The spatial distribution of the dust as well as the optical depth have a similar impact on $\Delta(\Psi)$. Therefore, the unambiguous determination of the density distribution requires complementary measurements, such as the measurement of the optical depth along the incident laser beam. For the analysis of the particle growth, however, it is sufficient to assume a simplified realization of the particle distribution and to compensate the discrepancy through the variation of the optical depth by scaling the amount of dust particles in the model. Ambiguities regarding the refractive index and the particle size distribution may also occur in the range of small particle radii and low optical depths. Since the refractive index has a significant impact on the derived particle sizes, it is important to resolve these ambiguities. The position of the maximum of $\Delta(\Psi)$ is a useful indicator to constrain the refractive index.

Based on our findings, we propose a strategy for in-situ diagnostics of the particle growth process, which we describe in detail in Sect.\,\ref{sec:diagnostic}. To keep the parameter space as small as possible, we propose to first fit only the refractive index and particle size distribution based on the assumption of single scattering and use it as a starting point for subsequent fitting to the entire $\Delta(\Psi)$ curve up to large particle radii. 

With the help of RT simulations and following the proposed strategy, we were able to analyze the growth characteristics of a particle growth experiment in a reactive argon acetylen plasma. In particular, we were able to reproduce the $\Delta(\Psi)$ curve up to a particle radius of about \SI{280}{\nano\meter}. Our best-fit model includes a change of the refractive index from $N = 2.051 + 0.075i$ to $N = 1.84 + 0.03i$ at a particle radius of about \SI{200}{\nano\meter}. This change is accompanied by a sudden increase of about \SI{20}{\nano\metre} in the particle growth rate. Before and after, the particles grow linearly with similar growth rates of \SI{1.96}{\nano\meter\per\second} and \SI{1.89}{\nano\meter\per\second}, respectively.
Interestingly, this growth behavior was also found in the course of a growth experiment, where dust particles were analyzed ex-situ \cite{dworschak_minimally_2021}. The method used based on the extraction of dust particles from the rim of the dust cloud. It is complementary to our in-situ approach and the results of both methods gives evidence that it is a global phenomenon of the whole dust cloud and not a local phenomenon at the spatial position of polarimeter. Sudden changes of the refractive index during the particle growth were also observed in a reactive a argon silan plasma \cite{hollenstein_1994_diagnostic}, i.\,e.\,the phenomenon is not a unique finding, but not investigated in detail. More detailed investigations, including more detailed ex-situ size analysis using both microscopy and imaging polarimetry (see below) should be done to clarify about the origin of the  occurring variation in the material properties of the growing particles.

RT simulations have proven to be a powerful tool for the in-situ analysis of dust clouds and, in particular, the growth process of the particles in the plasma. In the future, the importance of RT simulation will increase as the question of spatially resolved information about grain growth process arises. In particular, for the investigation of plasmas with inhomogeneous or layered particle growth imaging polarimetry is needed. For this purpose, the experimental setup will be expanded, e.g. with a laser sheet or even a full computer tomographic setup. \citeasnoun{greiner_2012_imie} and \citeasnoun{Groth_2019} present such "imaging Mie" setups, \citeasnoun{Groth_2019} shows that an imaging RCP has the ability to visualize spatiotemporally very inhomogeneous growth processes. The availability of polarizing cameras such as the Sonys Polarsens \cite{sony_polarsens} will significantly reduce the technical complexity of imaging polarimeters and demands for reliable data analysis methods.  Three-dimensional RT simulations are predestined to study the properties of imaging polarimeters in their very details.

\ack
We thank Florian Kirchschlager for helpful discussions and insight into his scripts for processing RT simulation results and Nils Rehbehn for valuable studies in the context of his bachelor and master thesis. We acknowledge support from the Deutsche Forschungsgemeinschaft DFG grant WO 857/19-1 and GR 1608/8-1.

\section*{References}
\bibliographystyle{jphysicsB}
\bibliography{PlasmaPaperI}

\end{document}